# Performance of RCPC-Encoded V-BLAST MIMO In Nakagami-*m* Fading Channel

Lydia Sari, Gunawan Wibisono, and Dadang Gunawan

**Abstract**— Multiple Input Multiple Output (MIMO) wireless communication link has been theoretically proven to be reliable and capable of achieving high capacity. However, these two advantageous characteristics tend to be addressed separately in many major researches. Researches on various approaches to attain both characteristics in a single MIMO system are still on-going and an established approach is yet to be concluded. To address this problem, in this paper a Vertical Bell Laboratories Layered Space-Time (V-BLAST) MIMO enhanced with Rate-Compatible Convolutional (RCPC) codes with Zero Forcing (ZF) and Minimum Mean Squared Error (MMSE)-based detection is proposed. The analytical BER of the system is presented and numerically analyzed. The system performance is analyzed in Nakagami-*m* fading channel, which provides accuracy and flexibility in matching the signals statistics compared to other fading models. The complexity which arises in the calculations of the RCPC codes parameters is significantly reduced by using equivalent convolutional codes. Results show that the use of high-rate code allows for bandwidth efficiency and at the same time does not severely degrades the system performance. It is also shown that the MMSE-based system outperforms the conventional ZF-based system especially in the low $E_b/N_0$ region and in severe fading conditions.

——————————— ◆ ———————————

## 1 INTRODUCTION

The need of high-capacity and highly-reliable wireless communication has been growing markedly over the last two decades. The multiple-input multiple-output (MIMO) scheme is one prominent candidate to fulfill this need, mainly due to its high capacity which does not call for bandwidth expansion [1-3]. The accomplishment of this high-capacity potential needs an appropriate signal processing architecture. Vertical Bell laboratories layered space-time (V-BLAST) is a MIMO architecture known to have spectral efficiencies of 20-40 bps/Hz at 24-34 dB average SNR without coding [2]. A suitable coding scheme is required to increase the reliability of V-BLAST MIMO.

In many wireless communication systems, convolutional codes are widely used due to their capability to enable soft-decision decoding and its near Shannon-capacity performance [4]. However the complexity in their decoding process tends to increase with each additional bit in the encoder output, and this may lead to erroneous decoding. One way to mitigate this problem is by using punctured convolutional codes, where several bits in the codewords produced by the encoders are deleted (punctured). The lesser number of bits in the codewords will make up for lesser branches in each node of the trellis, and therefore decreasing the system complexity. Punctured convolutional codes are later developed into Rate-Compatible Punctured Convolutional (RCPC) codes, in which code rates can be adapted to the protection levels of the transmitted data using single encoder and decoder structure [5].

Researches on MIMO systems with RCPC encoding are limited. One research on the effects of RCPC codes to complement a MIMO-OFDM scheme is [6]. The result shows that the RCPC-encoded MIMO-OFDM system has better performance compared to the same system without RCPC. However as no layered architecture is employed in [6], the spectral efficiency of the MIMO scheme is not fully exploited. The analytical BER for uncoded V-BLAST MIMO system with MMSE and ZF detection under Rayleigh fading condition has been presented in [7]. However the performance analysis in a more general fading model is still required especially for a coded V-BLAST MIMO system.

We therefore propose to use RCPC encoder to complement V-BLAST MIMO scheme, so as to exploit both the high-capacity nature of the system and to improve the system reliability. The fading model used in this research is Nakagami-*m*. This model provides more flexibility and accuracy in matching the signal statistics [4]. It can be used to model different fading conditions, including Rayleigh fading as a special case where *m* = 1.

The system performance will be affected by the detection criterion in the V-BLAST scheme, and the code performance. Several parameters that need to be calculated to measure the performance of punctured convolutional codes are

a) The minimum free distance of the code, $d_{free}$
b) The number of incorrect paths which distance $d \geq d_{free}$, diverging from the correct path to re-emerge with it at a later stage, $a_d$
c) The number of erroneous bits produced by the incorrect paths, $c_d$
d) The probability that an incorrect path is picked out during the Viterbi decoding process, $P_d$

The complexity involved in calculating the first three parameters for punctured convolutional codes will grow prohibitive as the number of shift registers in the encoder increases. Therefore a method to construct equivalent convolutional codes for punctured convolutional codes is needed to simplify the calculations. An equivalent convolutional code is used to represent a punctured convolutional code in its equivalent non-punctured form. Such representation is possible because punctured convolutional codes belong to a family of

————————————————

- *L. Sari is with the Department of Electrical Engineering, University of Indonesia, Depok 16424, Indonesia.*
- *G. Wibisono is with the Department of Electrical Engineering, University of Indonesia, Depok 16424, Indonesia.*
- *D. Gunawan is with the Department of Electrical Engineering, University of Indonesia, Depok 16424, Indonesia.*

linear codes and the resulting codeword shows periodic repeated pattern. Using the equivalent convolutional code, a punctured convolutional code can be analyzed more accurately in terms of its $d_{free}$, $c_d$ and $a_d$, compared to directly analyzing the trellis states of a punctured convolutional code for which no accurate methods are available.

The method to reconstruct an equivalent code which is valid for a general class of punctured convolutional code is reported in [8]. It does not however cover the code parameters attained through calculations using the equivalent codes.

Using the method based on [8] we reconstruct equivalent convolutional codes for RCPC codes with variable puncturing matrices to reach a code rate $R_{c2} = 2/3$ from a parent code with rate $R_{c1} = 2/4$. The performance parameters of these RCPC codes are calculated and tabulated for the various puncturing matrices used. The use of $R_{c2}$ implies bandwidth efficiency, as it uses shorter codewords compared to $R_{c1}$. However because it decreases the system performance, $R_{c1}$ is still needed especially in destructive channel condition. To identify the channel condition, the transmitter uses Channel State Information (CSI) that is fed back into it following a Singular Value Decomposition (SVD) process of the channel. A data stream entering a sub-channel with low level attenuation will be given low-rate code (high protection). On the contrary, a data stream entering a sub-channel with high level attenuation will be given high-rate code (low protection). This scheme ensures that extra bandwidth is not wasted to a data stream entering a destructive sub-channel.

The proposed RCPC-encoded V-BLAST MIMO system will use Zero Forcing (ZF) and Minimum Mean Squared Error (MMSE) criteria at the detection step. In the conventional ZF V-BLAST reported in [2], the ZF-based detection will inverse the channel gains. Using this detection, zeroes or very small eigenvalues in the MIMO channel will lead to noise enhancement and subsequently degrades the overall signal to noise ratio. This noise enhancement can be avoided by incorporating the noise term into the filtering process in the receiver. This method is known as MMSE-based detection, where the trade-off between noise enhancement and interference suppression occurs at the receiver. Therefore by using MMSE-based detection, the system performance is slightly improved [9-10].

This paper is organized as follows. Section 2 gives the system model covering the RCPC code design, the V-BLAST MIMO design with CSI and the system performance. Simulation results are discussed in Section 3 and the conclusion is given in Section 4.

## 2 SYSTEM MODEL
### 2.1 V-BLAST MIMO Model with CSI

The proposed system model is given Fig. 1. The number of transmit antennas is denoted as $M_t$ and the number of receive antennas is denoted as $M_r$.

CSI is fed onto the transmitter block, which in turn allocates the code rate to the transmit antennas, according to the condition of the sub-channel. The V-BLAST modulator in the transmitter block is basically a demultiplexer which maps a single stream of information bits onto the multiple transmit antennas. The elements of a signal vector $\mathbf{s} = [s_1, s_2,...,s_{Mt}]^T$ are transmitted simultaneously from first to the $M_t$-th transmit antennas, and the signal arriving at the receive antennas $\mathbf{y} = [y_1, y_2,...,y_{Mr}]^T$ can be expressed as

$$\mathbf{y} = \mathbf{Hs} + \mathbf{n} \qquad (1)$$

where $\mathbf{H}$ is the matrix channel of a MIMO system which elements are the channel gains between the transmit and receive antennas, and $\mathbf{n}$ is a noise vector with complex Gaussian distribution, zero mean and variance $\sigma_n$. To make the channel matrix known at the transmitter, it has to be decomposed into several single-input single-output (SISO) sub-channels. This can be done using Singular Value Decomposition (SVD) which decomposes the channel matrix into [3]

$$\mathbf{H} = \mathbf{U} \cdot \mathbf{D} \cdot \mathbf{V}^* \qquad (2)$$

where $\mathbf{U}$ and $\mathbf{V}$ are complex unitary matrices which dimensions are $M_r \times M_r$ and $M_t \times M_t$ respectively, $(.)^*$ denotes conjugate transpose and $\mathbf{D}$ is an $M_r \times M_t$ diagonal matrix which can be expressed as [3]

$$\mathbf{D} = \begin{bmatrix} \sqrt{\lambda_1} & 0 & \cdots & 0 \\ 0 & \sqrt{\lambda_2} & \cdots & 0 \\ \vdots & \vdots & \ddots & 0 \\ 0 & 0 & \cdots & \sqrt{\lambda_l} \end{bmatrix} \qquad (3)$$

where $\lambda_1,...,\lambda_l$ are the eigenvalues of $\mathbf{HH}^*$ and $l = \min(M_r,M_t)$. Taking the definition [3]

$$\mathbf{y} = \mathbf{U} \cdot \tilde{\mathbf{y}} \qquad (4.a)$$
$$\mathbf{s} = \mathbf{V} \cdot \tilde{\mathbf{s}} \qquad (4.b)$$
$$\mathbf{n} = \mathbf{U} \cdot \tilde{\mathbf{n}} \qquad (4.c)$$

and substituting (2) and (4) into (1), the received signal can be stated as

$$\tilde{\mathbf{y}} = \mathbf{D} \cdot \tilde{\mathbf{s}} + \tilde{\mathbf{n}} \qquad (5)$$

Equation (5) shows that the MIMO channel $\mathbf{H}$ has been decomposed into l equivalent parallel Singular-Input Singular-Output (SISO) sub-channels with the channel gains given by $\sqrt{\lambda_1},...,\sqrt{\lambda_l}$.

A special signal processing is needed at the receiver side to unmix the data streams. The V-BLAST demodulator extracts $\mathbf{s}$ from $\mathbf{y}$ using iterative nulling and cancellation process. The nulling and cancellation process is based on ZF and MMSE criteria. In the first step of the demodulation process, a decision statistic is used as a threshold to estimate the received signal. When ZF criterion is used for nulling, a matrix $\mathbf{w}_{ZF}$ which satisfies the following is needed

$$\mathbf{w}_{ZF} \left( \mathbf{U} \cdot \mathbf{D} \cdot \mathbf{V}^* \right) = \mathbf{I} \qquad (6)$$

with $\mathbf{I}$ as an identity matrix. The coefficient $\mathbf{w}_{ZF}$ is a Moore-Penrose pseudo-inverse of the channel matrix and is given by [11]

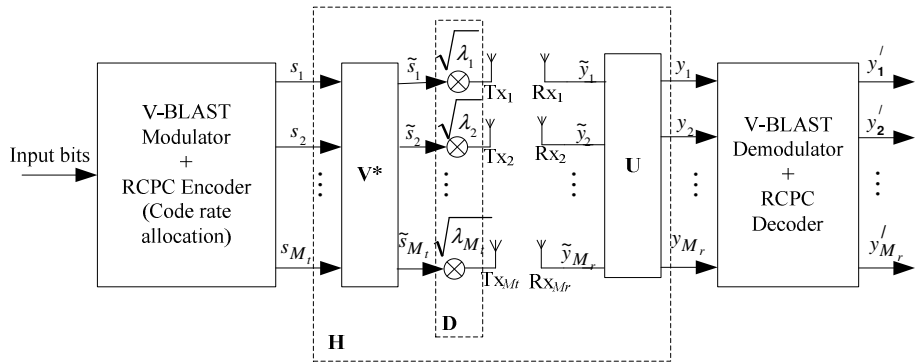

Fig. 1. System model of an RCPC-encoded V-BLAST MIMO

$$\mathbf{w}_{ZF} = (\mathbf{H}^*\mathbf{H})^{-1}\mathbf{H}^* \tag{7}$$

On the other hand, the MMSE approach requires a coefficient $\mathbf{w}_{MMSE}$ which minimizes the following

$$E((\mathbf{w}_{MMSE}\mathbf{y} - \mathbf{s})(\mathbf{w}_{MMSE}\mathbf{y} - \mathbf{s})^*) \tag{8}$$

Reference [11] solves the coefficient $\mathbf{w}_{MMSE}$ which gives

$$\mathbf{w}_{MMSE} = (\mathbf{H}^*\mathbf{H} + \sigma_n^2 \mathbf{I}_{M_t})^{-1}\mathbf{H}^* \tag{9}$$

where $\mathbf{I}_{Mt}$ denotes an $M_t \times M_t$ identity matrix. Using (7) and (9), the channel in (1) can be converted to identity matrix so that the transmitted signals can be estimated in the receiver. The extracted symbols in the receiver block is

$$\mathbf{y}' = \mathbf{w} \cdot \mathbf{y} \tag{10}$$

where $\mathbf{w}$ stands for either $\mathbf{w}_{ZF}$ or $\mathbf{w}_{MMSE}$ according to the criterion used. Using (10), the estimated symbols $\hat{s}$ can be obtained. The cancellation process done at the $i$-th step to extract a particular data stream from the mixed streams arriving at the receiver can be stated as

$$\mathbf{y}_i' = \mathbf{y} - \sum_{j=1}^{i-1} \mathbf{h}_j \hat{s}_j \qquad i = 1,...,M_t \tag{11}$$

where $h_j$ denotes the $i$-th column of $\mathbf{H}$.

The system performance is the probability that the signal estimated at the receiver does not match the transmitted signal. This is a function of the system modulation, fading channel, and the performance of error correction codes used and will be given in Section 2.3.

To evaluate the system performance, the post processing SNR per bit for the MIMO system needs to be defined. In [7], the post-processing SNR per bit for V-BLAST MIMO is defined using QR decomposition of $\mathbf{H}$ so that

$$\mathbf{H} = \mathbf{QR} \tag{12}$$

where $\mathbf{R}$ is an $M_t \times M_t$ upper triangular matrix and $\mathbf{Q}$ is an $M_r \times M_t$ matrix which orthonormal columns are the ZF nulling vectors. The matrix $\mathbf{R}$ can be stated as

$$\mathbf{R} = \begin{bmatrix} R_{11} & R_{12} & \ldots & R_{1N} \\ 0 & R_{22} & \ldots & R_{2N} \\ \vdots & \ddots & \ddots & \vdots \\ 0 & \ldots & 0 & R_{M_tM_t} \end{bmatrix} \tag{13}$$

The elements of $\mathbf{R}$ stand for the channel gains and will be used for defining post-processing SNR per bit for both ZF and MMSE V-BLAST MIMO systems.

**2.2 RCPC Codes**

RCPC codes are developed from punctured convolutional codes. In RCPC codes a rate-compatibility is implemented to ensure that high-rate codes are embedded to the low-rate codes of the same family [5]. Fig. 2 illustrates an RCPC encoder with mother code rate $R_{c1} = k/n = 2/4$, and the number of shift register used is S = 5. The puncturing period is $P_c = 2$. A zero in the puncturing matrix indicates a punctured bit. In Fig. 2, all input bits passing through Encoder 0 will be transmitted. However, one bit out of the 4 output bits entering Encoder 1 will be deleted, as denoted by a zero in the puncturing matrix. The use of puncturing matrices decrease the number of transmitted bits from 2 ☐ $P_c$ to $P_c + \delta$ per $P_c = 2$ information bits, where $1 \leq \delta \leq (n-1) P_c$. In this example, as $n = 2$ then the code rates generated may take the value of 2/3 or 2/4. Therefore in this example $R_{c1} = 2/4$ and $R_{c2} = 2/3$.

The puncturing matrix is a binary matrix which dimension is $n \times P_c$, where $n$ is the number of bits of a codeword. A convolutional code which rate is $1/n$ can be stated as a $K/nK$ code, with $K$ being any value. The code with rate $K/nK$ is called $K$ times blocked code [8] and the primary $1/n$ code is called the mother code.

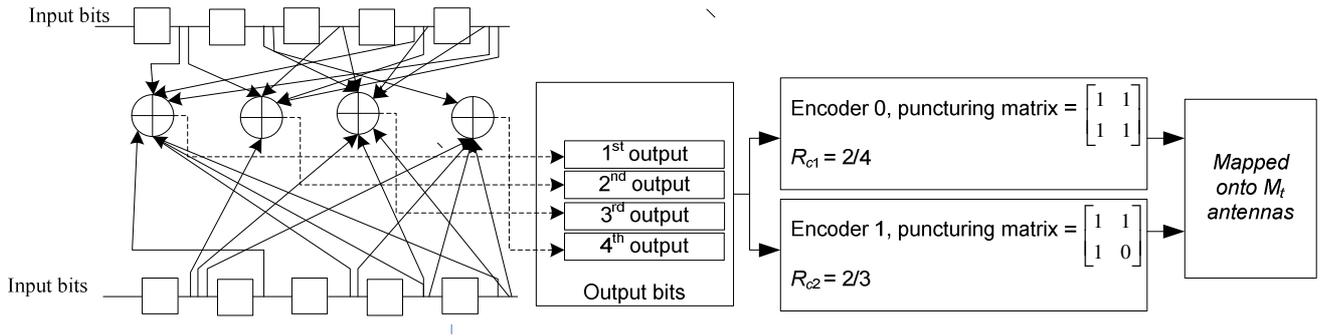

Fig. 2. Example of an RCPC Encoder with Rc1 =2/4 and Rc2 = 2/3

A code with generator matrix $G = (G_0, \ldots, G_{n-1})$ can be expanded into a blocked code which generator matrix $G'$ consists of $nK$ polynomials $P_{ij}$, which satisfy the following [8]

$$G_i = \sum_{j=0}^{K-1} D^j P_{ij}(D^K) \qquad i = 0, \ldots, n-1 \qquad (14)$$

where $D^j$ and $D^K$ are the $D$-transform of the generator matrix with respect to $i$ and $K$, respectively. Each of the $n$ polynomials defining $G_i$ is further split into $K$ polynomials such that [8]

$$G' = \begin{cases} P_{j \bmod n, \lfloor j/n \rfloor - i} & \text{if } n \times i \leq j \\ DP_{j \bmod n, \lfloor j/n \rfloor - i + K} & \text{if } n \times i > j \end{cases} \qquad (15)$$

where $\lfloor x \rfloor$ denotes the smallest integer closest to $x$. For example, if a mother code of rate $R_{c1} = 1/2$ is blocked 2 times ($K = 2$), the equivalent rate is $R_{c1} = 2/4$. The expanded matrix $G'$ consists of $K = 2$ elements, where each element is further split into 2 polynomials $P_{ij}$, to yield $(P_{00}, P_{01})$, $(P_{10}, P_{11})$. From (14), however, it is known that the elements of the expanded generator may include $DP_{00}$, $DP_{01}$, $DP_{10}$ and $DP_{11}$. Using (15), the exact elements of the equivalent generator matrix can be figured out. The next step is to arrange the elements into a complete expanded generator matrix which can be stated as [8]

$$G' = [Z^{K-1} \times Mat \mid Z^{K-2} \times Mat \mid \ldots \mid Z \times Mat \mid Mat] \qquad (16)$$

where $Z$ is a $K \times K$ matrix consisting of an upper diagonal of 1, a $D$ in the bottom left corner and a 0 elsewhere. The expressions stated in (15) and (16) complement each other and cannot be used independently to define $G'$. The matrix $Mat$ which dimension is $K \times n$ consists of polynomials $P_{ij}$. For $K = 2$ and $n = 2$ and the encoders depicted in Fig. 2, $G'$ can be stated as

$$G' = [Z \times Mat \mid Mat]$$
$$= \begin{bmatrix} \begin{bmatrix} 0 & 1 \\ D & 0 \end{bmatrix} \begin{bmatrix} P_{01} & P_{11} \\ P_{00} & P_{10} \end{bmatrix} \mid \begin{bmatrix} P_{01} & P_{11} \\ P_{00} & P_{10} \end{bmatrix} \end{bmatrix} \qquad (17)$$
$$= \begin{bmatrix} P_{00} & P_{10} & P_{01} & P_{11} \\ DP_{01} & DP_{11} & P_{00} & P_{10} \end{bmatrix}$$

To reconstruct the punctured convolutional codes, the punctured columns are deleted from the equivalent convolutional codes. Using the puncturing matrix used for Encoder 1 in Fig. 2 which is

$$\begin{bmatrix} 1 & 1 \\ 1 & 0 \end{bmatrix} \qquad (18)$$

where the fourth column of the coded bits is punctured, the resulting equivalent punctured convolutional code is

$$G' = \begin{bmatrix} P_{00} & P_{10} & P_{01} \\ DP_{01} & DP_{11} & P_{00} \end{bmatrix} \qquad (19)$$

There are no systematic methods available to construct good rate $P_c/(P_c+\delta)$ codes [5]. Consequently, the search for good codes in this paper is based on the best-known generator polynomial of rate 1/2. This generator polynomial forms the mother code and blocked $K$ times to yield the equivalent code.

In this research $R_{c1} = 2/4$ which is punctured with $P_c = 2$ to yield $R_{c2} = 2/3$ is used. $R_{c2}$ is attained using various puncturing matrices, and the resulting $c_d$ values are shown in Table 1. The high level protection given by $R_{c1}$ will be applied to the information stream entering sub-channels with low level attenuation, as opposed to $R_{c2}$ which will be applied to the stream entering sub-channels with high level attenuation. This arrangement ensures that the high level protection is given only to the data entering a non-destructive sub-channel. RCPC codes has a probability of error that can be stated as [5]

$$P_{b,RCPC} \leq \frac{1}{P_c} \sum_{d=d_{free}}^{\infty} c_d P_d \qquad (20)$$

The parameter $P_d$ is a function of the modulation and fading channel types as well as the decoding error and will be stated in the next section.

### 2.3 System Performance

The modulation used for this research is square $M$-QAM where $M$ denotes the constellation size of the signal, $M = 2^c$ points ($c$ even). A maximum ratio combining (MRC) scheme is applied at the receiver side, and must be considered in the system performance calculation.

The conditional BER for a square $M$-QAM signal in a multichannel MRC reception is [4],[12]

$$P_{b,M-QAM}(\gamma) = \frac{4}{k}\left(1 - \frac{1}{\sqrt{M}}\right) Q(\sqrt{2g\gamma}) - \frac{4}{k}\left(1 - \frac{1}{\sqrt{M}}\right)^2 Q^2(\sqrt{2g\gamma}) \qquad (21)$$

TABLE 1 PERFORMANCE PARAMETERS FROM MOTHER CODE WITH CODE GENERATOR [657 435] YIELDING $R_{C1} = 2/4$ AND $R_{C2} = 2/3$

| Encoder | $d_{free}$ | 5 | 6 | 7 | 8 | 9 | 10 | 11 | 12 | 13 | 14 | 15 |
|---|---|---|---|---|---|---|---|---|---|---|---|---|
| 0 | $\begin{bmatrix}1 & 1\\1 & 1\end{bmatrix}$ (no puncturing) | | | | 3 | 0 | 47 | 0 | 263 | 0 | 2017 | 0 |
| 1 | $\begin{bmatrix}0 & 1\\1 & 1\end{bmatrix}$ | 8 | 26 | 8 | 263 | 1470 | 5272 | 21705 | 99998 | 424070 | 1747352 | 7265287 |
| 2 | $\begin{bmatrix}1 & 0\\1 & 1\end{bmatrix}$ $c_d$ | | 9 | 17 | 181 | 774 | 3140 | 13737 | 60959 | 262324 | 1094392 | 4613797 |
| 3 | $\begin{bmatrix}1 & 1\\0 & 1\end{bmatrix}$ | 9 | 10 | 58 | 400 | 1968 | 8575 | 35003 | 149563 | 637000 | 2649633 | 10935387 |
| 4 | $\begin{bmatrix}1 & 1\\1 & 0\end{bmatrix}$ | catastrophic | | | | | | | | | | |

where

$$g = \frac{3}{2(M-1)} \quad (22)$$

and the Q-functions take the alternate forms given in [12]

$$Q(x) = \frac{1}{\pi} \int_0^{\pi/2} \exp\left[\frac{-x^2}{2\sin^2\theta}\right] d\theta, \quad x \geq 0 \quad (23)$$

$$Q^2(x) = \frac{1}{\pi} \int_0^{\pi/4} \exp\left[\frac{-x^2}{2\sin^2\theta}\right] d\theta, \quad x \geq 0 \quad (24)$$

The total post-processing SNR per bit $\gamma$ can be stated as

$$\gamma = \sum_{l=1}^{M_t M_r} A_l^2 \frac{cE_b}{N_0} \quad (25)$$

where $A$ denotes the fading amplitude. Using (22) – (24) the average BER can be stated as

$$P_{b,M-QAM} = \frac{4}{k\pi}\left(1 - \frac{1}{\sqrt{M}}\right) \int_0^{\pi/2} \prod_{l=1}^{M_r M_t} I_l(\gamma_l, g, \theta) d\theta$$
$$- \frac{4}{k\pi}\left(1 - \frac{1}{\sqrt{M}}\right)^2 \int_0^{\pi/4} \prod_{l=1}^{M_r M_t} I_l(\gamma_l, g, \theta) d\theta \quad (26)$$

where $I_l(\gamma_l, g, \theta)$ denotes a function of modulation and fading channel type and tabulated in [12]. For a Nakagami-$m$ channel, this function is defined as [12]

$$I_l(\gamma_l, g, \theta) = \left(1 + \frac{g\bar{\gamma}}{m\sin^2\theta}\right)^{-m} \quad (27)$$

Using the same steps used in [12] for the case of correlated fading channel, the average BER of a square $M$-QAM system with MRC can be stated as

$$P_{b,cor} = \frac{4}{c\pi}\left(1 - \frac{1}{\sqrt{C}}\right) \int_0^{\pi/2} Cor(g, L, \gamma, \rho, \theta) d\theta$$
$$- \frac{4}{c\pi}\left(1 - \frac{1}{\sqrt{C}}\right)^2 \int_0^{\pi/4} Cor(g, L, \gamma, \rho, \theta) d\theta \quad (28)$$

where $L$ denotes the number of signals arriving at the receiver, $L = M_r \times M_t$ and the correlated fading model is [12]

$$Cor(g, L, \gamma, \rho, \theta) = \left(1 + \frac{\gamma \cdot g \cdot r}{m \cdot L \cdot \sin^2\theta}\right)^{-mL^2/r} \quad (29)$$

where the correlation factor is $0 \leq \rho \leq 1$ and

$$r = L + \frac{2\sqrt{\rho}}{1-\sqrt{\rho}}\left(L - \frac{1-\left(\sqrt{\rho}\right)^L}{1-\left(\sqrt{\rho}\right)}\right) \quad (30)$$

The post processing $E_b/N_0$ for ZF V-BLAST system is [7]

$$\left(\frac{E_b}{N_0}\right)_{ZF} = R_{ii}^2 \left(\frac{E_b}{N_0}\right)_{in}, \quad i = 1,...,N \quad (31)$$

while for MMSE V-BLAST it is defined as

$$\left(\frac{E_b}{N_0}\right)_{MMSE} = \left(\frac{E_b}{N_0}\right)_{ZF} + \eta_{MMSE} \quad (32)$$

The parameter $\eta_{MMSE}$ is defined in [7] as

$$\eta_{MMSE} = \beta_i^* \cdot \left(\mathbf{H}_{i-1}^* \mathbf{H}_{i-1} + \sigma_n^2 \mathbf{I}_N\right)^{-1} \beta_i \quad (33)$$

where $\beta_i$ is an ($i$-1) dimension circularly symmetric Gaussian vector.

The performance of uncoded V-BLAST MIMO systems can be analyzed using (26) and (28), and applying the appropriate post-processing $E_b/N_0$ according to the chosen detection criterion.

To analyze the coded V-BLAST MIMO system the parameter $P_d$ in (20) needs to be defined. For an RCPC-encoded multi-channel system, the upper-bound for $P_d$ is

$$P_d = \int_{-\infty}^{\infty} P(d,\alpha) f_\alpha(\alpha) d\alpha \tag{34}$$

where $d = d_{free}, \ldots, \infty$ and $\alpha$ is a random variable defined as

$$\alpha = \sum_{l=1}^{d} \left(\frac{E_s}{N_0}\right)_l = \sum_{l=1}^{d} \left(\frac{c \cdot R_c \cdot E_b}{N_0}\right)_l \tag{35}$$

and the parameter $P(d,\alpha)$ can be defined as

$$P(d,\alpha) = Q\left[\sqrt{\frac{D_{min}^2}{2E_s} \cdot \alpha}\right] \tag{36}$$

Where $D_{min}$ is the minimum Euclidean distance of the modulation constellation. To define $f_\alpha(\alpha)$ for a Nakagami-$m$ channel, first we state the probability density function (pdf) for Nakagami-$m$ channel,

$$f_z(z) = \frac{2m^m}{\Gamma(m)} z^{2m-1} \exp(-mz^2) \tag{37}$$

Substituting z for a random variable $\left(\frac{E_s}{N_0}\right)_l$ and following the steps in [13] yields

$$f_{\left(\frac{E_s}{N_0}\right)_l}\left(\frac{E_s}{N_0}\right)_l = \frac{m^m}{\Gamma(m)\left[\left(\overline{E}_s/\overline{N}_0\right)_l\right]^m}\left[\left(\frac{E_s}{N_0}\right)_l\right]^{m-1} \exp\left(\frac{-m\left(\frac{E_s}{N_0}\right)_l}{\left(\overline{E}_s/\overline{N}_0\right)_l}\right) \tag{38}$$

Therefore the pdf of $\alpha$ is

$$f_{\alpha l}(\alpha) = \frac{m^{md}}{\Gamma(md)\left[\left(\overline{E}_s/\overline{N}_0\right)_l\right]^{md}} \alpha^{md-1} \exp\left(\frac{-m\alpha}{\left(\overline{E}_s/\overline{N}_0\right)_l}\right) \tag{39}$$

The parameter $\left(\overline{E}_s/\overline{N}_0\right)_l$ can be stated as

$$\left(\overline{E}_s/\overline{N}_0\right)_l = c \cdot R_c \cdot \left(\frac{\overline{E}_b}{\overline{N}_0}\right) \tag{40}$$

Substituting (36), (39) and (40) into (34) yields

$$P_d = \int_{-\infty}^{\infty} Q\left(\frac{D_{min}^2 \alpha}{2E_s}\right) \frac{m^{md}}{\Gamma(md)\left[\left(cR_c\frac{\overline{E}_b}{\overline{N}_0}\right)\right]^{md}} \alpha^{md-1} \exp\left(\frac{-m\alpha}{\left(cR_c\frac{\overline{E}_b}{\overline{N}_0}\right)}\right) \tag{41}$$

The pdf of V-BLAST MIMO system in Nakagami-$m$ fading channel has to be taken into account. To model the pdf of such system, the joint pdf of two signals, each undergoing Nakagami-$m$ fading as stated in (37) is used. The result is

$$f_\xi(\xi) = \frac{2m^{2m}}{\Gamma(m)^2} \xi^{4m-1} \exp(-m\xi^2) \tag{42}$$

It is observed that $m$ as a fading parameter in Nakagami-$m$ fading is related to the degree of freedom $t$ in Chi distribution. As (42) belongs to a central Chi-square distribution, which pdf is

$$f_\omega(\omega) = \frac{1}{\Gamma\left(\frac{t}{2}\right)} z^{\frac{t}{2}-1} \exp\left(-\frac{\omega^2}{2}\right) \tag{43}$$

and following [14] the degree of freedom of the system is twice the diversity order, comparing (42) to (43) yields

$$t = 8m \tag{44}$$

where $t$ can be defined as [14]

$$t = 2(M_r + M_t - 1) \tag{45}$$

Subtituting (45) to (44) yields

$$m = \frac{(M_r + M_t - 1)}{4} \tag{46}$$

Using (20), (26), (42) the probability of error for an RCPC-encoded V-BLAST MIMO system in Nakagami-$m$ independent fading channel is

$$P_b = P_{b,M-QAM}\left[\frac{2\left(\frac{|M_r - M_t|+1}{4}\right)^{\frac{|M_r-M_t|+1}{2}}}{\Gamma\left(\frac{M-N+1}{4}\right)^2}\right]\overline{\frac{E_b}{N_0}}^{|M_r-M_t|}$$

$$\times \exp\left(-\frac{|M_r - M_t|+1}{4}\left(\frac{\overline{E}_b}{\overline{N}_0}\right)^2\right)\frac{1}{P_c}\sum_{d=d_{free}}^{\infty} c_d P_d \tag{47}$$

whereas for correlated Nakagami-$m$ channel the system performance as yielded by (20), (28), (42) is

$$P_b = P_{b,cor}\left[\frac{2\left(\frac{|M_r - M_t|+1}{4}\right)^{\frac{|M_r-M_t|+1}{2}}}{\Gamma\left(\frac{|M_r - M_t|+1}{4}\right)^2}\right]\overline{\frac{E_b}{N_0}}^{|M_r-M_t|}$$

$$\times \exp\left(-\frac{|M_r - M_t|+1}{4}\left(\frac{\overline{E}_b}{\overline{N}_0}\right)^2\right)\frac{1}{P_c}\sum_{d=d_{free}}^{\infty} c_d P_d \tag{48}$$

In the final two equations the post-processing SNR per bit used is either $\left(\frac{E_b}{N_0}\right)_{ZF}$ or $\left(\frac{E_b}{N_0}\right)_{MMSE}$ according to the detection criterion used. Leaving out (42) from the probability of error formula will yield the probability of error for MRC system, and further changing $M_r = M_t = 1$ in the MRC system will yield the performance of a SISO system.

## 3 RESULTS AND DISCUSSIONS

The system performance for ZF V-BLAST MIMO system with $M_r = M_t = 2$, and $P_c = 2$ under independent and correlated Nakagami-$m$ fading where $m = 0.5$ is depicted in Fig. 3. The code rates used are $R_{c1} = 2/4$ and $R_{c2} = 2/3$ as given in Table 1. The correlation coefficient used is $\rho = 0.8$. It is shown that the use of punctured codes slightly lowers the system performance. However this penalty is not severe as the punctured codes allows the transmission of shorter codeword for each uncoded bit, which in turn contributes to bandwidth efficiency. It is also shown that the different puncturing matrices used to generate codes with $R_{c2} = 2/3$ yield similar performances. Performance difference is noted at the low $E_b/N_0$ region, and tends to converge in high $E_b/N_0$ region. This is due to the fact that the code performances yielded by the different puncturing patterns are similar. Although from Table 1 it is noted that Encoder 2 has the largest $d_{free}$ value, it only differs by 1 from the other encoders' $d_{free}$ values. It is shown that the signals undergoing highly correlated fading environment has slight performance degradation compared to the signals undergoing independent fading.

The performance of the same system under $m = 3$ and 5 is given in Fig. 4. As correlated fading condition does not exceptionally affect the performance of the proposed system, in Fig. 4 only signals under independent fading are numerically simulated. It is shown that under the better fading figures, the system performance improves by approximately 1 dB.

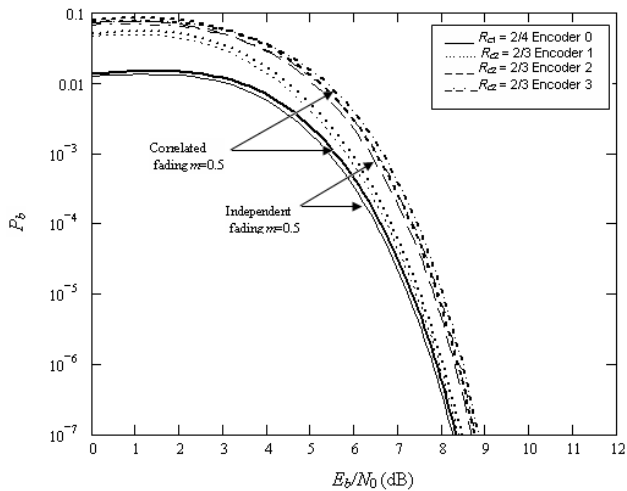

Fig. 3 System performance of ZF V-BLAST MIMO with $M_r = 2$, $M_t = 2$

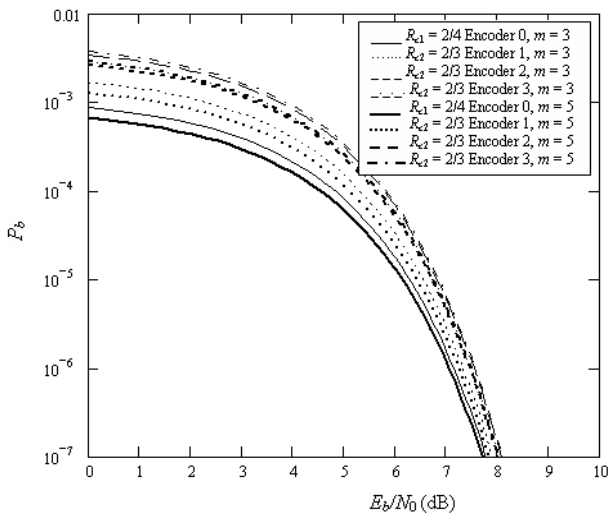

Fig. 4. System performance of ZF V-BLAST MIMO with $M_r = 2$, $M_t = 2$

The same configuration $M_r = M_t = 2$ with $P_c = 2$ and MMSE detection criterion is used for the system depicted in Fig. 5. Compared to the system using ZF detection criterion depicted in Fig. 3, the use of MMSE criterion improves the system performance by approximately 0.5 dB. It is shown that the use of MMSE detection criterion will improve the system performance more noticeably in the low $E_b/N_0$ region. The use of MMSE criterion also sets the performances of the signals undergoing independent and correlated fading further apart compared to the ZF criterion. As the MMSE criterion holds noise error into account, the use of punctured codes will yield a more noticeable performance degradation compared to ZF criterion. However, the degradation caused by puncturing is still less than 0.5 dB for both ZF and MMSE criteria.

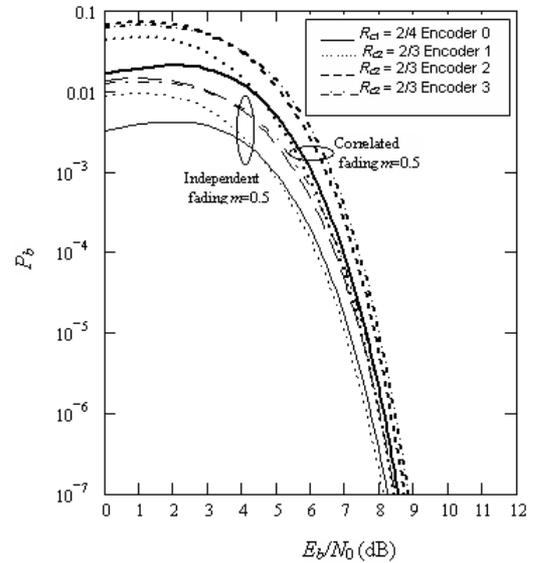

Fig. 5. System performance of MMSE V-BLAST MIMO with $M_r = 2$, $M_t = 2$

Using the same configuration, $M_r = M_t = 2$ and MMSE detection criterion, the system performance under $m = 3$ and 5 is given in Figure 6. Again the performance improvement is approximately 0.5 dB at high $E_b/N_0$ region.

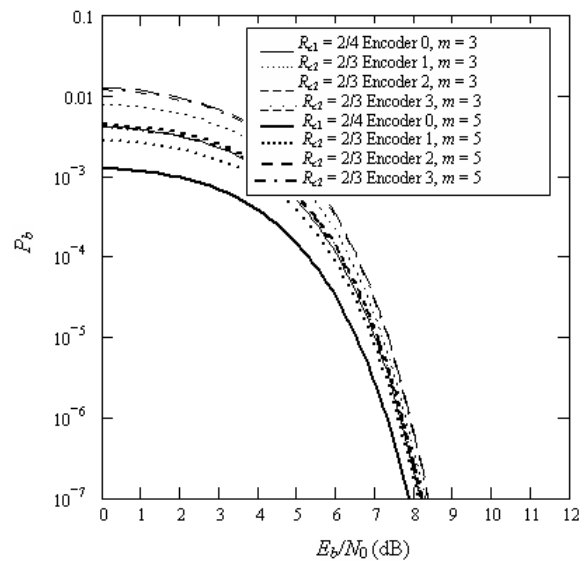

Fig. 6. System performance of MMSE V-BLAST MIMO with $M_r = 2$, $M_t = 2$

The performance of a system where $M_r = 4$, $M_t = 2$, $m = 0.5$ using ZF criterion is given in Fig. 7. Compared to the system with $M_r = M_t = 2$ depicted in Fig. 3, the increased number of receive antennas improves the system performance by approximately 3 dB. Fig. 8 shows the system where ZF criterion is used, when $M_r = 2$ and $M_t = 4$. Compared to the system where $M_r = M_t = 2$, the system performance improves by approximately 2 dB. However in the low $E_b/N_0$ region the system performance is worse. This result suggests that the increase of receive antennas will contribute more to the system performance improvement, as opposed to increasing the number of transmit antennas.

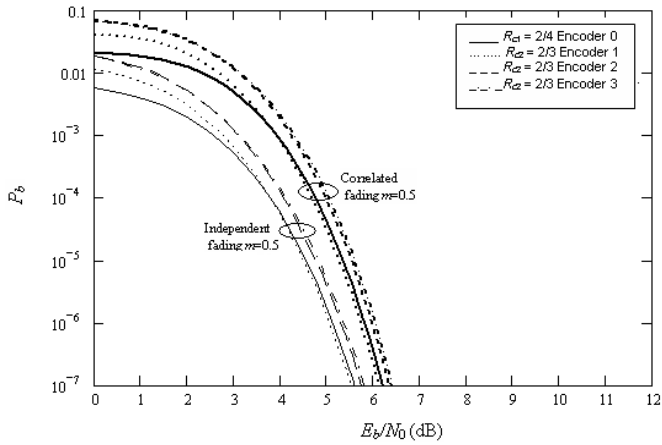

Fig. 7. System performance of ZF V-BLAST MIMO with $M_r = 4$, $M_t = 2$

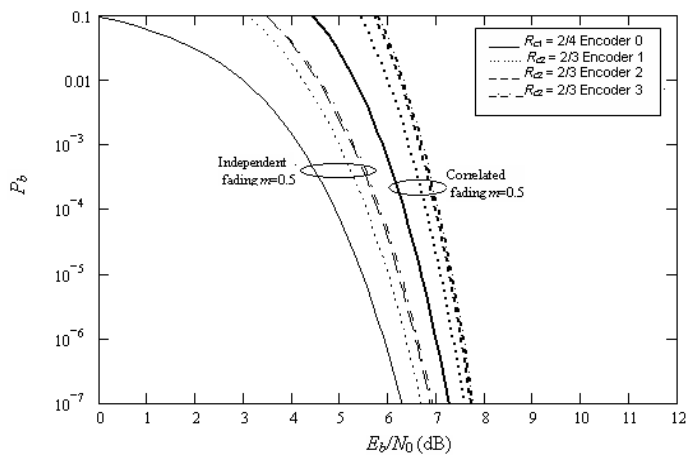

Fig. 8. System performance of ZF V-BLAST MIMO with $M_r = 2$, $M_t = 4$

The performance of a system which uses MMSE detection criterion with $M_r = 4$, $M_t = 2$ and $m = 0.5$ is given in Fig. 9. Compared to the system depicted in Fig. 5, the performance improved by approximately 3 dB. Compared to the system using ZF with the same antenna configuration, it is observed that MMSE criterion yields a better performance in the low $E_b/N_0$ region for independent and correlated fading conditions.

When MMSE criterion is used for a system with $M_r = 2$ and $M_t = 4$ as depicted in Fig. 10, a performance improvement of approximately 2 dB is obtained compared to the same system with $M_r = 2$ and $M_t = 2$. Compared to the system which uses ZF criterion, $M_r = 4$ and $M_t = 2$, it is shown that the MMSE criterion allows for a smaller $E_b/N_0$ range for independent and correlated fading conditions. Comparing all antenna configurations and detection criteria, it is observed that MMSE V-BLAST system with 2 transmit and 4 receive antennas yields the best system performance. The system performance is most prominent in the low $E_b/N_0$ region.

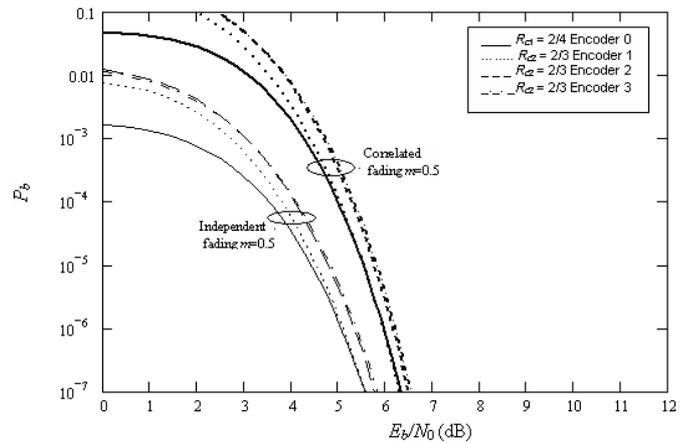

Fig. 9. System performance of MMSE V-BLAST MIMO with $M_r = 4$, $M_t = 2$

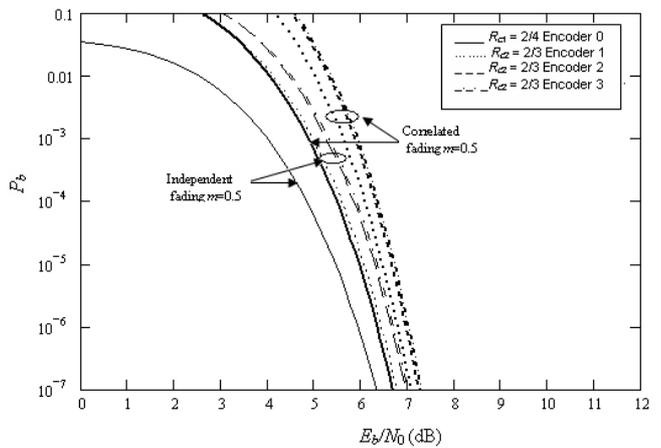

Fig. 10. System performance of MMSE V-BLAST MIMO with $M_r = 2$, $M_t = 4$

## 4 CONCLUSION

A V-BLAST MIMO system enhanced with RCPC codes and both ZF and MMSE-based detection has been proposed and analyzed. The mathematical BER for the system under Nakagami-$m$ fading channel condition is presented, as this particular fading type can represent more general fading models. Numerical simulation results show that the use of RCPC codes to complement a V-BLAST MIMO system increases the robustness of the system, even in severe fading condition. The high- and low-rate codes attained by using RCPC are used to encode signals entering destructive and non-destructive sub-channels respectively, to allow for bandwidth efficiency. It is shown that the bandwidth efficiency does not come with the cost of great performance degradation, as the use of punctured codes decreases the system performance by less than 1 dB. It is also shown that the MMSE-based detection yields better performance compared to the ZF-based detection of the proposed system, especially in the low $E_b/N_0$ region. The increase of antenna numbers also contributes to better system performance. However, the performance improvement is greater if the number of receive antennas is increased, as opposed to the increasing the number of transmit antennas. The best performance is given by MMSE-based V-BLAST MIMO system with 2 transmit and 4 receive antennas with the most prominent improvement observed in the low $E_b/N_0$ region. Further researches are needed to clarify the significance of puncturing patterns to increase the free distance of RCPC codes and subsequently improve the BER performance of the proposed system.

**Lydia Sari** received the graduate and post-graduate degrees in electrical engineering from Trisakti University, Indonesia and University of Indonesia, Indonesia, in 1998 and 2002 respectively. She is now working towards the PhD degree in Electrical Engineering Department, University of Indonesia, Indonesia.

**Gunawan Wibisono** received the B.Sc. degree in electrical engineering from University of Indonesia in 1990, and M.Eng and Ph.D degrees from Keio University, Japan, in 1995 and 1998, respectively. He is a lecturer in Electrical Engineering Department, University of Indonesia. His research interests are wireless communication and coding theory. Dr. Wibisono is a member of IEEE and Indonesian Telecommunication Society.

**Dadang Gunawan** received the B.Sc. degree in electrical engineering from University of Indonesia in 1983, and M.Eng and Ph.D degrees from Keio University, Japan, and Tasmania University, Australia in 1989 and 1995, respectively. He is the Head of Telecommunication Laboratory and the Wireless and Signal Processing Research Group of the Electrical Engineering Department, University of Indonesia. His research interests are wireless communication and signal processing. Dr. Gunawan is a senior member of IEEE.